\documentstyle[epsfig]{mn}
\begin{document}
\def\ltsima{$\; \buildrel < \over \sim \;$}
\def\simlt{\lower.5ex\hbox{\ltsima}}
\def\gtsima{$\; \buildrel > \over \sim \;$}
\def\simgt{\lower.5ex\hbox{\gtsima}}
\def\approxgt{\mathrel{\hbox{\rlap{\lower.55ex \hbox {$\sim$}}
        \kern-.3em \raise.4ex \hbox{$>$}}}}
\def\approxlt{\mathrel{\hbox{\rlap{\lower.55ex \hbox {$\sim$}}
        \kern-.3em \raise.4ex \hbox{$<$}}}}

\title[Hint of a cyclotron emission feature in IGR~J18483$-$0311]{{\itshape XMM-Newton} and INTEGRAL study of the SFXT IGR~J18483$-$0311 in quiescence: hint of a cyclotron emission feature?}

\author[Sguera et al.]
{V. Sguera$^{1,2}$, L. Ducci$^{3,4}$, L. Sidoli$^3$, A. Bazzano$^1$, L. Bassani$^2$
\\
$^1$ INAF, Istituto di Astrofisica Spaziale e Fisica Cosmica, Via Fosso del Cavaliere 100, I-00133 Rome, Italy\\
$^2$ INAF, Istituto di Astrofisica Spaziale e Fisica Cosmica, Via Gobetti 101, I-40129 Bologna, Italy \\
$^3$ INAF, Istituto di Astrofisica Spaziale e Fisica Cosmica, Via E. Bassini  15, I-20133 Milano, Italy \\
$^4$ Dipartimento di Fisica e Matematica, Universit\`a degli Studi  dell'Insubria, Via Valleggio 11, I-22100 Como, Italy \\
}

\date{Accepted 2009 December 4. Received in original form 2009 November 6}

\maketitle

\begin{abstract}
We report the results from archival {\itshape XMM-Newton} and INTEGRAL observations of the Supergiant Fast X-ray Transient (SFXT)
IGR~J18483$-$0311 in quiescence. The 18--60 keV hard X-ray behaviour of the source is presented here for the first time, 
it is characterized by a spectral shape ($\Gamma$ $\sim$2.5) similar to that during outburst activity and the 
lowest measured luminosity level is $\sim$10$^{34}$ erg   s$^{-1}$. The 0.5--10 keV luminosity state, measured by {\itshape XMM-Newton}
during  the apastron passage, is about one order of magnitude lower and it is reasonably fitted  by an absorbed black body model yielding
parameters consistent with previous measurements. In addition, we
find evidence ($\sim$3.5$\sigma$ significance) of an emission-like feature at $\sim$3.3 keV in the quiescent 
0.5--10 keV source spectrum. The absence of any known or found systematic effects, which could artificially introduce the observed feature,
give us confidence about its non-instrumental nature. We show that its physical explanation 
in terms of atomic emission line appears unlikely and conversely we attempt to ascribe 
it to an electron cyclotron emission line which would imply  
a neutron star magnetic field of the order of $\sim$3$\times$10$^{11}$ G. Importantly, such direct estimation 
is in very good agreement with that independently inferred by us in the framework of 
accretion from  a spherically symmetric stellar wind. If firmly confirmed by future 
longer X-ray observations, this would be the first detection ever of a cyclotron feature in the X-ray spectrum of a SFXT, with
important implications on theoretical models.

\end{abstract}

\begin{keywords}
X-rays:binaries -- X-rays: individual (IGR J18483$-$0311)
\end{keywords}

\vspace{1.0cm}

\section{Introduction}
IGR~J18483$-$0311 is a hard X--ray transient
discovered in outburst 
with INTEGRAL in 2003 (Chernyakova et al. 2003).
Other hard X-ray outbursts were observed showing 
fluxes and durations  up to $\sim$120 mCrab (20--100 keV) and $\sim$2 days, respectively (Sguera et al. 2007). 
The X--ray emission shows two periodicities: a longer one at $\sim$18.55 days 
which is interpreted as orbital  (Levine et al. 2006, Sguera et al. 2007), and a shorter one at $\sim$21.0526 s which 
was discovered with the X--ray monitor JEM-X (Sguera et al. 2007).
Giunta et al. (2009) performed a study of the quiescent soft X-ray behaviour of the source (0.5--10 keV)
and  confirmed the pulse period 
through {\itshape XMM-Newton} data, providing a 
measurement of the spin-period derivative which is very likely due to light travel time effects in the binary system. 
A complete monitoring of the X--ray emission from IGR~J18483$-$0311 over an entire orbital period 
has been performed in 2009 with Swift/XRT (Romano et al. 2009).
The optical counterpart has been identified with a B0.5Ia supergiant star located at a distance of 3--4 kpc 
(Rahoui et al. 2008). This, together with the transient nature of the source, implies its classification 
as a Supergiant Fast X--ray Transient  (SFXTs; Sguera et al. 2005, 2006, Negueruela et al. 2005, Sidoli 2009).
\section{{\itshape XMM-Newton} observation and results}
\subsection{Data reduction} 
We analyzed an archival {\itshape XMM-Newton} EPIC observation of IGR J18483$-$0311 
collected on 2006 October 12 for a total exposure of $\sim$19 ks. However, after 
rejecting time intervals  affected by high background the net good exposure time reduced to 14.4 ks. 
{\itshape XMM-Newton} data were reprocessed using version 9.0 of the Science Analysis
Software (SAS). Known hot, or flickering, pixels and electronic
noise were rejected using the SAS. 
The ancillary and response matrices
were generated using the SAS tasks {\em arfgen} and {\em rmfgen}.
Spectra were selected
from single and double events only (pattern from 0 to 4) for the EPIC pn 
full frame mode while for
both MOS cameras patterns from 0 to 12 were selected.
Source counts were extracted from circular regions
of 40 arcseconds radius centered on the source for both pn and  MOS.
Background counts were obtained from similar
regions offset from the source position. The backgrounds do not
show any evidence for flaring activity, so that the entire nominal
exposure times were considered. 
The source net count rates (1--10 keV) are the following:
0.097$\pm{0.003}$ counts~s$^{-1}$ (pn), 0.035$\pm{0.002}$ counts~s$^{-1}$ (MOS1), and 0.027$\pm{0.001}$ counts~s$^{-1}$ (MOS2).
The source is too faint for a meaningful spectral analysis with the RGS.
Also the spectroscopy with the MOS cameras does not provide any improvement with
respect to the analysis of  the pn spectrum alone, thus in the following 
we will concentrate only
on the source EPIC pn results.
In order to ensure applicability of the $\chi^{2}$ statistic, the
extracted spectra were rebinned such that at least 20 counts per
bin were present and such that the energy resolution was not
over-sampled by more than a factor of 3. 
The spectral analysis  was performed using Xspec version 12.5; 
all quoted uncertainties are given at the 90\% confidence level
for one single parameter of interest.

\subsection{EPIC-pn spectral analysis} 
We first fit the 0.5--10 keV  EPIC-pn spectrum of IGR J18483$-$0311  with an absorbed power law model
whose best fit parameters (N$_{H}$=7.8$^{+1.3}_{-1.1}$$\times$10$^{22}$ cm$^{-2}$, $\Gamma$=2.4$\pm$0.3) are in fair agreement with 
those already reported by Giunta et al. (2009). However, we note that such model gives a statistically inadeguate representation of  the observed spectrum being the $\chi^{2}_{\nu}$=1.46 for 52 d.o.f. (see also Giunta et al. 2009 where 
$\chi^{2}_{\nu}$=1.3 for 39 d.o.f.).  This motivates us to investigate  alternative spectral models.
An absorbed  thermal bremsstrahlung is still a bad statistical description of the data 
($\chi^{2}_{\nu}$=1.34, 52 d.o.f.), on the contrary an absorbed thermal black body results in a much better 
fit ($\chi^{2}_{\nu}$=1.17, 52 d.o.f.) and yields parameters (N$_{H}$=3.4$^{+0.6}_{-0.5}$$\times$10$^{22}$ cm$^{-2}$, kT=1.35$\pm$0.08 keV)
consistent  with previous measurements by Romano et al. (2009) from a Swift/XRT monitoring campaign 
covering an entire orbital period. The upper and middle panels in Fig. 1 show the absorbed black body fit 
spectrum and the ratio of data to model, respectively. 
It is worth noting that such spectral fit resulted  in a radius of the emitting black body region
equal to $\sim$0.14 km, i.e. consistent with a small portion of the neutron star surface such as its polar cap region.  
The unabsorbed (observed) 0.5--10 keV flux is 1.24$\times$10$^{-12}$ erg  cm$^{-2}$ s$^{-1}$ (9.1$\times$10$^{-13}$ erg  cm$^{-2}$ s$^{-1}$) which translates into a X-ray luminosity of 1.3$\times$10$^{33}$ erg   s$^{-1}$ by assuming a distance of the optical counterpart of 3 kpc. 
If we follow Sguera et al. (2007) and measure the source phase from the epoch of the brightest outburst observed 
with INTEGRAL at MJD 53844.2 (phase 0), then
the {\itshape XMM-Newton} observation took place at orbital phase 0.52 (i.e. right 
during the apastron passage) and the above X-ray luminosity value represents the lowest quiescent X-ray state of the source 
ever reported in the literature. For the sake of clarity, here and in the following we refer to ``quiescence'' 
as to the lowest  detectable level of X-ray activity from the source (i.e. $\sim$10$^{33}$ erg   s$^{-1}$) during which it is still 
accreting matter even if outside its bright outbursts activity (i.e. $\sim$10$^{36}$ erg   s$^{-1}$).

\begin{figure}
\centering
{\includegraphics[height=8.0cm,width=4.5cm,angle=-90]{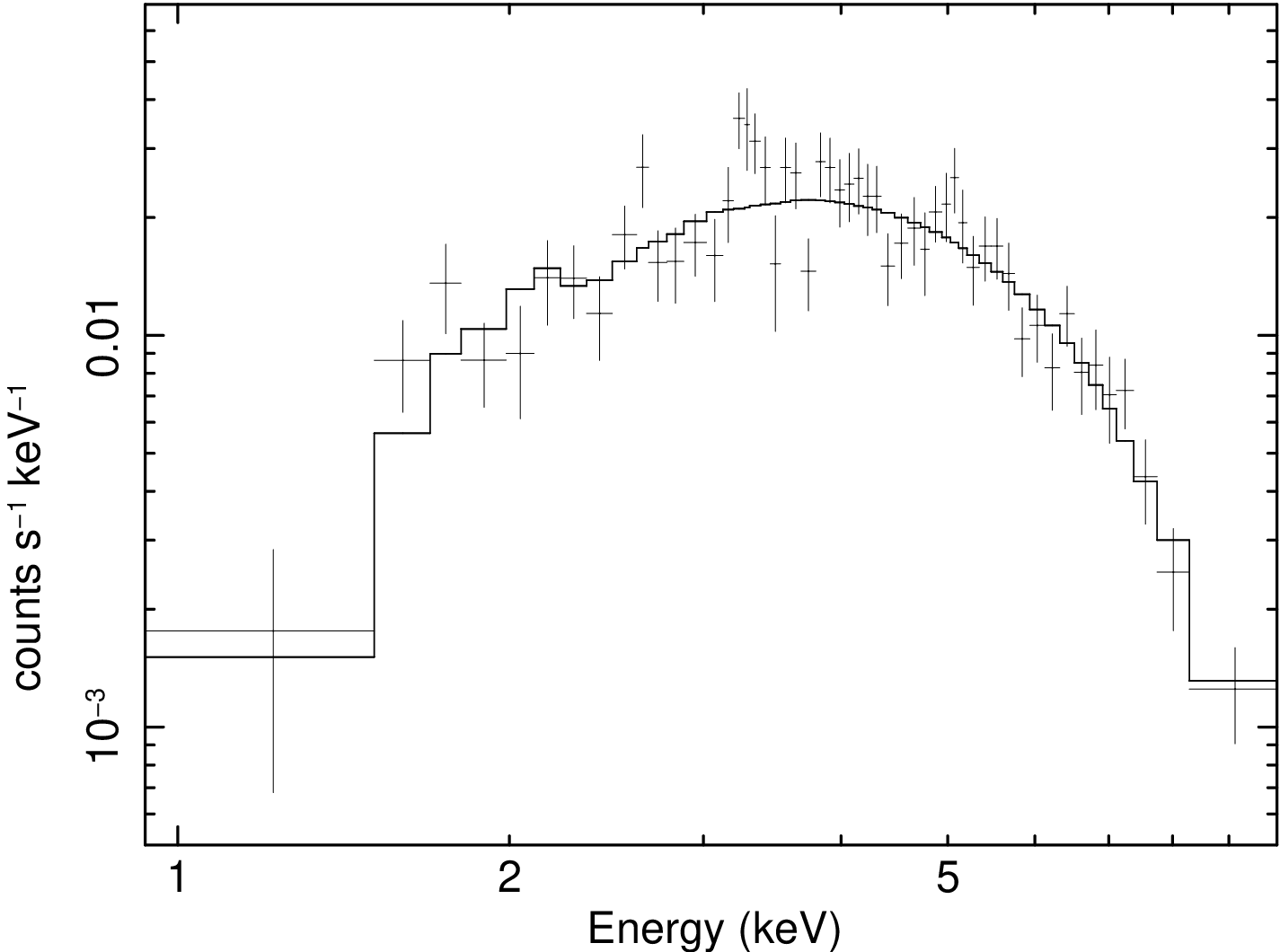}} \quad
{\includegraphics[height=8.0cm,width=2.0cm,angle=-90]{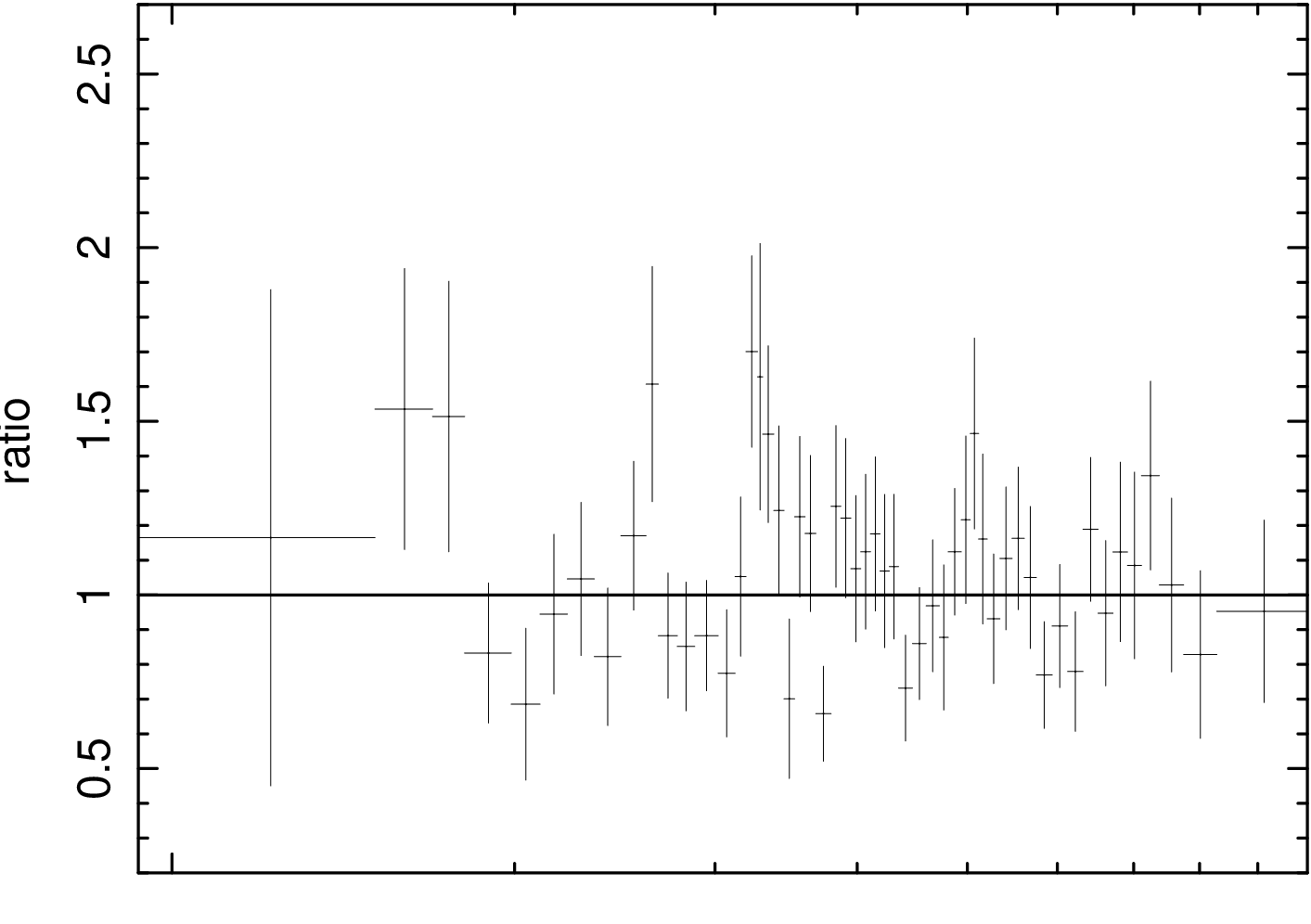}} \\
{\includegraphics[height=8.0cm,width=2.0cm,angle=-90]{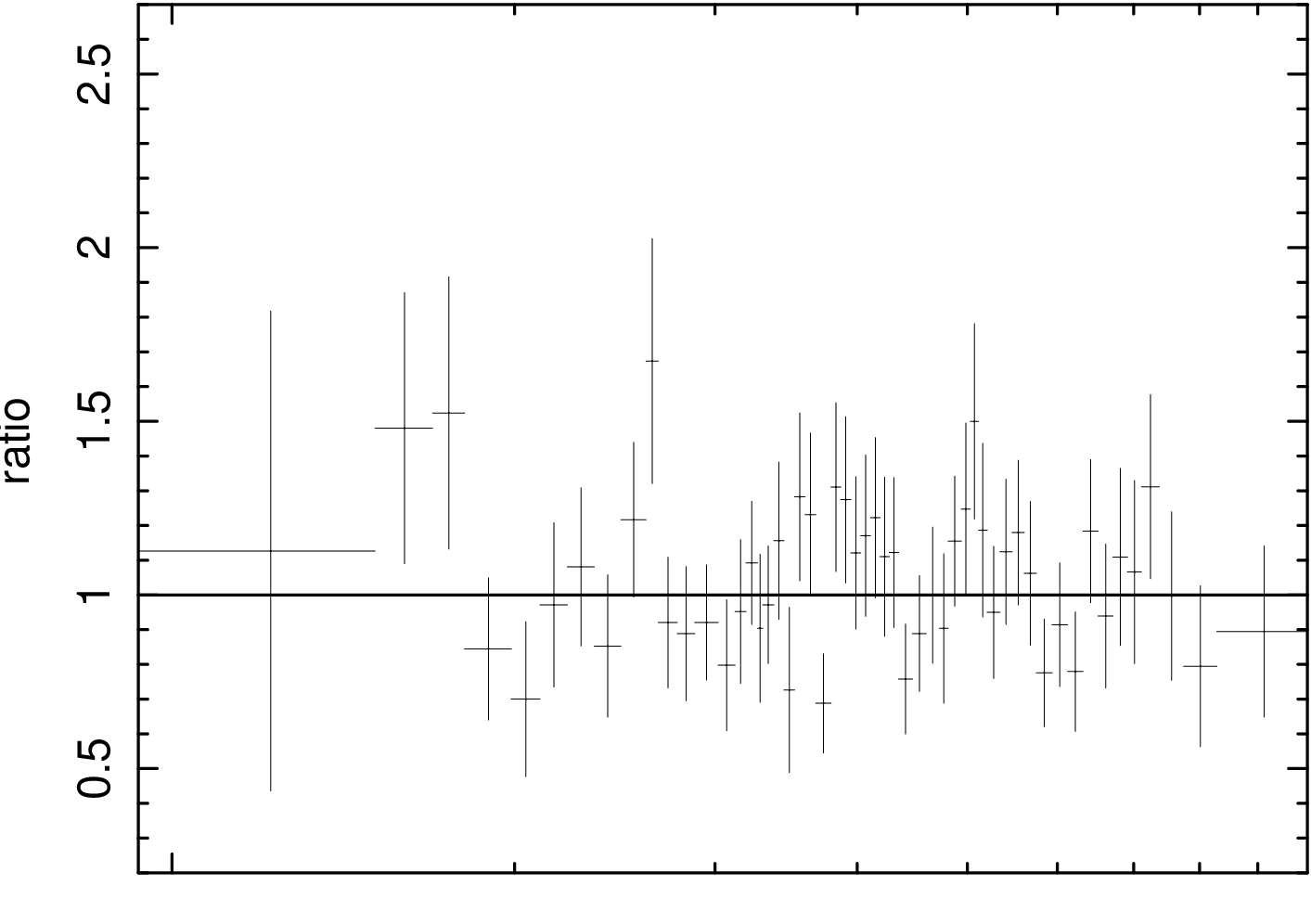}} \quad
\caption{\emph{Upper  panel}: data-to-model (absorbed black body) of the 0.5--10 keV EPIC-pn spectrum of IGR J18483$-$0311. 
\emph{Middle panel}: the ratio of data to model is shown for the fit with no gaussian line included in the model. 
\emph{Lower panel}: the ratio of data to model is shown again but this time including a gaussian line.}
\end{figure}



\subsubsection{An emission-like feature at $\sim$3.3 keV?}
As stated in the previous section, the absorbed black body model is a reasonable statistical description
of the continuum; however the ratio of data to model clearly show an excess around $\sim$3.3 
keV (see Fig. 1, middle panel), suggesting the presence of a 
possible spectral line. To model such residuals we added  a gaussian emission line with energy, width and normalisation free to vary.
We obtained a significant improvement of the statistical quality of the fit ($\chi^{2}_{\nu}$=0.97, 49 d.o.f) corresponding 
to a significance level of confidence of 99.4\% according to a simple F-test ($\sim$3$\sigma$ significance). 
We are aware that the F-test is not a good measure of the actual significance of such  
additional spectral line feature (see Protassov et al. 2002), however it could give an indication. To this aim, we point out that  
the obtained low F-test probability value should make the detection of the line stable against 
mistakes in the calculation of its significance. Furthermore, we note that there are in the literature a few 
cases of cyclotron features reported with a F-test probability value very similar to our one (e.g. Rea et al. 2003, Ibrahim et al. 2002).
Even using an absorbed power law model, as already adopted by Giunta et al. (2009), 
we still have evidence of the line although at lower statistical significance (96.7\%). 
We measured an energy centroid of 3.28$^{+0.046}_{-0.041}$ keV; the line is unresolved and  it has an equivalent width 
of 148$^{+226}_{-80}$ eV. The luminosity of the line is $\sim$2.1$\times$10$^{31}$ erg   s$^{-1}$, i.e. $\sim$2\% of the total
0.5--10 keV luminosity. 
The robustness of the line detection is attested by the energy centroid-normalisation parameters confidence contours 
shown in Fig. 2; moreover the line intensity is equal to 5.5$^{+1.5}_{-1.7}$ $\times$10$^{-6}$ photon cm$^{-2}$ s$^{-1}$ for a significance detection of $\sim$3.5$\sigma$. Other positive excesses in the ratio of data to model are present
in the spectrum at $\sim$2.5 keV and $\sim$5 keV (see Fig. 1, middle panel), but they are not significative.

In order to rule out  an artificial nature of the spectral emission 
line due to instrumental effects,  we checked that the response matrix did not introduce any strong feature around 3.3 keV 
(i.e no sharp variation is present in the spectral response). 
The inspection of the background spectrum did not reveal any spectral feature at the same energy. 
We also note that no uncalibrated instrumental features or edges  are expected or known  
close to $\sim$3.3 keV.  The absence of any known or found  systematic effects 
give us confidence about the non-instrumental nature of the observed spectral line. 
In the following  we discuss its possible physical origin. 

Firstly, we took into account the possibility that the observed emission feature could be due to atomic 
transition lines. In fact,  if originating in an accretion gas or corona or in the circumstellar material, atomic lines could well be
present in the X-ray spectra of accreting pulsar and they are expected to be mainly narrow and in emission.
According to the atomic database ATOMDB\footnote{http://cxc.harvard.edu/atomdb/WebGUIDE/index.html}, we found that two highly ionized Argon atomic lines
are expected close to $\sim$3.3 keV (i.e. Ar XVIII at 3.32 keV and  Ar XVII at 3.14 keV). However, we consider unlikely a physical explanation 
in terms of Ar atomic lines because of the following reasons:
i) it would require an unexplainable high overabundance of Ar which is not obvious in the neutron star atmosphere or in its surrounding ambient, 
ii)  it would not explain why only the highly ionized Ar line is observed and no other lines are seen from 
 more abundant elements, iii) it would not justify the lack of a similar spectral feature in any of the many known accreting X-ray pulsar in HMXBs
for which much higher signal to noise X-ray spectra are available. Therefore, we explored an alternative physical explanation in terms of  
cyclotron line. In fact, depending on the physical condition of  the emitting/absorbing material in the accretion column, 
cyclotron features from electrons or protons could be observed in the X-ray spectra of accreting magnetized pulsars (Heindl et al. 2004). 
They provide an important tool for a direct measurement of the neutron star magnetic field value, since the  observed fundamental energy
is related to the magnetic field value by E$^{obs}_{c,e}$=11.6(B/10$^{12}$ G) in the case of electrons or by 
E$^{obs}_{c,p}$=0.63(B/10$^{14}$G) in the case of protons.  If interpreted as an electron cyclotron 
emission feature,  the putative line observed  at $\sim$3.3 keV implies a magnetic field 
of $\sim$2.8$\times$10$^{11}$ G (if the forming region is situated 
far above the pulsar's polar cap) or  alternatively of $\sim$3.7$\times$10$^{11}$ G (if it is situated close to the neutron star surface, 
i.e. at the base of the accretion column,  and so affected by a gravitational redshift of z=0.3) .
For protons cyclotron feature, the magnetic field should be even higher by a factor of $\sim$1836 (the proton to electron mass ratio) implying a value of 
$\sim$5$\times$10$^{14}$ G (i.e a magnetar-type magnetic field).

\begin{figure}
\begin{center}
\epsfig{file =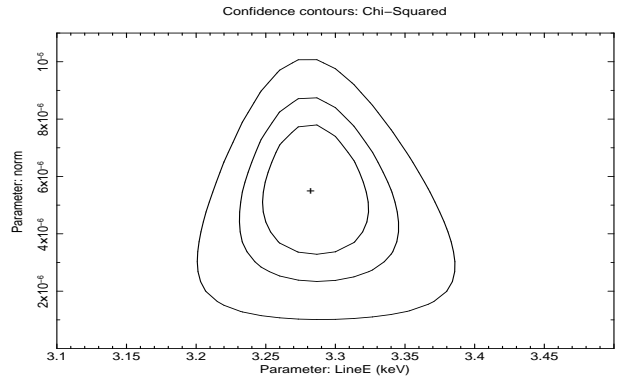,height=8cm,width=5cm, angle=-90}
    \caption{Confidence contours level (68\%,95\%,99\%) for the energy line center and normalization parameter values.} 
\end{center}
\end{figure}

\section{INTEGRAL observations  and results} 
The quiescent behaviour of  IGR J18483$-$0311 in the soft X-ray band (0.5--10 keV) is fairly known and studied
(Romano et al. 2009, Giunta et al. 2009) while above 20 keV is totally unknown. 
With the aim of investigating for the first time the hard X-ray quiescent state of the source,
we collected INTEGRAL data with IBIS (Ubertini et al. 2003).
INTEGRAL observations are typically divided into short pointings called Science Windows (ScWs) each of $\sim$2000 s duration. 
The data reduction was carried out with the release 7.0 of the Offline Scientific Analysis (OSA) software.

Firstly, we searched the entire IBIS/ISGRI  public data archive for pointings where IGR J18483$-$0311 was 
within the fully coded field of view  of ISGRI (9$^{\circ}\times$9$^{\circ}$). 
Subsequently, since our aim was to measure the quiescent hard X-ray emission,
we intentionally excluded those individual ScWs during which the source was in outburst, i.e. significantly detected at 
$\geq$5$\sigma$. By doing so, we collected a total of 401 ScWs which were used to generate a mosaic 
significance map in the 18--60 keV band for a total exposure 
of $\sim$ 795 ks. IGR J18483$-$0311 was weakly detected at $\sim$7$\sigma$ level with an average flux (luminosity) of 1.5 mCrab 
(1.3$\times$10$^{34}$ erg  s$^{-1}$), likely representing the lowest quiescent hard X-ray state of the source.  The relative spectrum  
(18--60 keV) is  equally well fit using a power law ($\Gamma$=2.5$\pm$0.8, $\chi^{2}_{\nu}$=0.8, 15 d.o.f.) 
or alternatively a black body model (kT=7$\pm$2 keV, $\chi^{2}_{\nu}=0.8$, 15 d.o.f.): such spectral shape is 
similar to that seen during the outburst activity (Sguera et al. 2007). 

A joint spectral fit to the {\itshape XMM-Newton} and IBIS/ISGRI quiescent data, although non simultaneous, was also attempted. 
A multiplicative factor for each instrument was included in the fit to take into account the 
uncertainty  in the cross calibration of the instruments. 
We used spectral models usually adopted to describe the X-ray emission from accreting pulsars.
Firstly we found that the 0.5--60 keV broad band spectrum can be adequately described ($\chi^{2}_{\nu}$=0.95, 64 d.o.f.)
by a phenomenological model such as absorbed power law together with a black body at low energies, to describe the continuum, 
plus a gaussian line in emission at 3.3 keV (we left all parameters free to vary except for the absorption N$_H$ which was fixed 
to its nominal value previously found from the {\itshape XMM-Newton} spectral analysis). The best fit parameters are $\Gamma$=2.0$^{+0.5}_{-0.6}$,
kT=1.40$^{+0.06}_{-0.09}$ and E$_{c}$=3.28$\pm$0.05.  Subsequently, we used a more physical thermal Comptonization model, namely the Bulk Motion Comptonization (Titarchuk et al. 1997).
The spectral parameters of such model are the black body colour temperature kT$_{bb}$, the spectral index 
$\alpha$ which is related to the Comptonization efficiency (when $\alpha$ is smaller then the 
efficiency of energy transfer from the hot electrons to the soft seed photons is higher) and finally logA (where 
A is the illuminating factor) which gives an indication of the fraction 
of the up-scattered black body photons with respect to the black body seed photons directly visible.
The BMC model reproduced rather well the data ($\chi^{2}_{\nu}$=0.93, 64 d.o.f.) both at high and low energies 
where it was only modified by photoelectric absorption and a gaussian line (see Fig. 4.). The best fit parameters are kT$_{bb}$=1.40$\pm$0.07 keV, 
$\alpha$=1.37$\pm$0.9 and log(A)=--1.04$^{+1}_{-0.8}$.  
We note that although the intercalibration costant between {\itshape XMM-Newton} and IBIS is expected to be around 1, we found a 
smaller value. 
In principle this could imply some variability in the source flux between the
observations, however the most reasonable explanation is that the soft X-ray flux measured by {\itshape XMM-Newton} at apastron passage 
represents the lowest  quiescent state of the source while the hard X-ray flux, measured by IBIS 
over a much longer time interval, represents the average quiescent state along the eccentric orbit.

\begin{figure}
\begin{center}
\epsfig{file =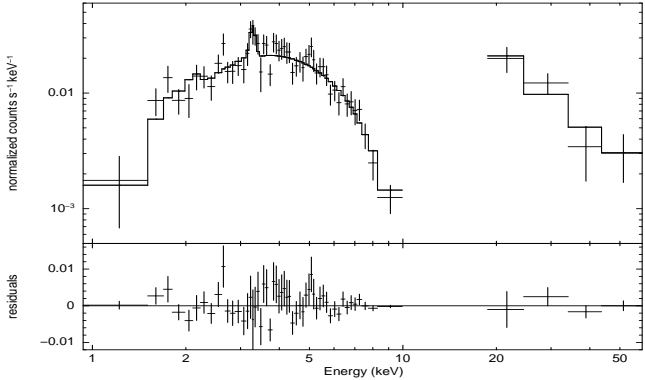,height=8.5cm,width=5cm, angle=-90}
    \caption{Broad-band spectrum (0.5--60 keV) of IGR J18483$-$0311 fitted with a Bulk Motion Comptonization model modified 
at low energies by photoelectric absorption and a gaussian line. The lower panel shows the residuals from the fit.} 
\end{center}
\end{figure}

\section{Alternative estimation of the neutron star magnetic field}


As previously reported in section 2 and 3, IGR J18483$-$0311 has been detected in quiescence by  
{\itshape XMM-Newton} and IBIS   with a luminosity in the range 10$^{33}$--10$^{34}$ erg s$^{-1}$ and  spectral shape similar to
that during outburst activity. Such information could be used within the framework of 
accretion from  a spherically symmetric stellar wind to obtain an alternative and independent estimation of the 
magnetic field. In fact, the material in the background wind will be accreted onto
the neutron star only if the magnetospheric radius $R_{\rm M}$
is less than the corotation radius $R_{\Omega}$ ($R_\Omega > R_{\rm M}$),
otherwise centrifugal forces will expel the matter 
(Davidson \& Ostriker 1973; Stella et al. 1983).
We are aware that the above assumed model and propeller effect are based on 
various and stronlgy simplified assumptions, however this is still a reasonable approach to obtain  a rough and reliable estimate
for the magnetic field strength  and the eccentricity.
The magnetospheric radius is obtained by balancing the matter pressure
to the magnetic field pressure, i.e.
$\rho V^2 = B(R_{\rm M}^2)/8\pi$ where $B(R_{\rm M}) = B_{\rm NS} R_{\rm NS}^3/R_{\rm M}^3$ ($R_{\rm NS}$ is the radius of the neutron star,
$M_{\rm NS}$ is the mass and $B_{\rm NS}$ its surface magnetic field).
Although the measured $M_{\rm NS}$ in accreting X-ray systems is known 
to be roughly in the range 1--2.5  M$_\odot$,  neutron stars masses determinated at the highest accuracy display a normal distribution centered at the 
value of 1.35 M$_\odot$ with a very small dispersion of $\pm$0.04 M$_\odot$ (Thorsett \& Chakrabarty 1999). 
Bearing this in mind, here and in the following calculations 
we assume the mass of the neutron star as a costant equal to its canonical value 
of $1.4$~M$_\odot$. The corotation radius is given by:
\begin{equation}
R_\Omega = \left ( \frac{G M_{\rm NS} P^2}{4\pi^2} \right )^{1/3} \mbox{.}
\end{equation}
where P is the spin period. Assuming a typical neutron star mass of $1.4$~M$_\odot$
and P=$21.05$~s, as measured for IGR~J18483$-$0311,
we get
R$_\Omega$=1.27$\times$10$^9$ $\mbox{ cm.}$

The observed quiescent X-ray luminosities 
could be produced only if the background wind material can reach the surface of the neutron star
and impact onto it.
However, the requirement of $R_\Omega > R_{\rm M}$
imposes important constraints.

We calculated the Bondi-Hoyle X-ray luminosity 
at orbital phases $\phi=0$ and $\phi=0.52$
assuming a supergiant mass $M_{\rm OB}=33$~M$_\odot$
and a radius $R_{\rm OB}=33.8$~R$_\odot$ (Searle et al. 2008),
and $M_{\rm NS}=1.4$~M$_\odot$, $R_{\rm NS}=10$~km for the neutron star.
Then we assumed the typical wind parameters
for a star with same spectral type
of the IGR~J18483$-$0311 supergiant:
a mass loss rate $\dot{M}$ in the range 
$0.4-3 \times 10^{-6}$~M$_\odot$~yr$^{-1}$,
the wind velocity law $v(r)=v_\infty(1-R_{\rm OB}/r)^\beta$, 
where $v_\infty = 1200-1900$~km~s$^{-1}$, 
$\beta = 0.8 -1.6$
and we leave the magnetic field (B) and the orbital eccentricity (e) as 
free parameters. 
We calculated the expected X-ray luminosity of the neutron star 
at periastron and at the orbital phase $\phi \sim 0.52$, 
assuming that the X-ray emission is due 
to accretion of matter from the spherically symmetric 
stellar wind emitted by the supergiant companion.
The neutron star accretes only the fraction of the wind
ejected by the OB star within a distance from 
the neutron star smaller than:
\begin{equation} \label{accretion radius}
R_{\rm a} = \frac{2 G M_{\rm NS}}{v_{\rm rel}^2 + c_{\rm s}^2}
\end{equation}
here $R_{\rm a}$ is called \emph{accretion radius}, and $v_{\rm rel}$
is given by:
\begin{equation} \label{v_rel}
v_{\rm rel}^2(r) = (v_{\rm wind}(r) - v_{\rm r}(r))^2 + v_{\phi}^2(r)
\end{equation}
where $v_{\rm rel}$ is the relative velocity of the wind material $v_{\rm wind}$ with
respect to the radial $v_{\rm r}$ and tangential $v_{\phi}$ 
components of the neutron star orbital velocity,
and $c_{\rm s}$ is the sound velocity.
The fraction of stellar wind captured by the neutron star
is given by the mass flux which goes through the area $\pi R_{\rm a}^2$,
and is given by:
\begin{equation} \label{M_accr}
\dot{M}_{\rm acc}= \rho_{\rm wind}(r)v_{\rm rel}(r) \pi R_{\rm a}^2
\end{equation}
where $\rho_{\rm wind}(r)$ is the wind density
related to the rate of mass loss by the stellar wind ($\dot{M}$)
by means of the continuity equation.
From Equations (\ref{accretion radius}), (\ref{v_rel}), (\ref{M_accr}),
we computed the X-ray luminosity $L_{\rm x}$ of a neutron star:
\begin{equation} \label{L_x}
L_{\rm x, wind} = \frac{G M_{\rm NS}}{R_{\rm NS}}\dot{M}_{\rm acc} = \frac{(G M_{\rm NS})^3}{R_{\rm NS}} \frac{ 4 \pi \rho_{\rm w}(r)}{[(v_{\rm rel}^2(r) + c_{\rm s}^2]^{3/2} }
\end{equation}
where $G$ is the gravitational constant.
%
%
The allowed range of parameters 
(e and B) which reproduce the X-ray luminosity 
at the periastron ($\phi \sim 0$) and apastron ($\phi \sim 0.52$), calculated according to Equation (\ref{L_x}), 
are  B$\la$6$\times$10$^{11}$ G and  0.39$\la$e$\la$0.62.  We note that 
the above estimated magnetic field is in very good agreement
with the value independently inferred by us from the putative electron cyclotron line at 
$\sim$3.3 keV ($\sim$3--4$\times$10$^{11}$ G).

\section{Summary and conclusions}
In this work we studied the SFXT IGR J18483$-$0311 in quiescence using archival INTEGRAL and {\itshape XMM-Newton} data.
As results, we report for the first time the  spectral properties ($\Gamma$=2.5$\pm$0.8) 
and luminosity measurement (1.32$\times$10$^{34}$ erg  s$^{-1}$) of the average quiescent 
hard X-ray emission above 18 keV, accumulated over a long time interval along the eccentric orbit. 
On the contrary, the 0.5--10 keV  observation took place in coincidence with the apastron passage and allowed 
the measurements of the lowest quiescent  state of the source (1.3$\times$10$^{33}$ erg   s$^{-1}$); 
its spectrum is best fitted by an absorbed black body with  parameters consistent
with those previously reported by Romano et al. (2009) from a Swift/XRT monitoring campaign.

One important point deserves further and deeper attention. An intriguing spectral 
emission line has been observed  at $\sim$3.3 keV in the {\itshape XMM-Newton} quiescent spectrum of IGR J18483$-$0311 at apastron passage and we speculated about its physical origin. An explanation in terms of atomic emission lines (i.e. highly ionized Argon) appears unlikely. 
We attempt to ascribe it  to an  electron cyclotron emission 
line from an X-ray pulsar accreting at a low rate, as predicted by Nelson et al. (1995,1993).
Specifically, these authors predicted the possible detection   
of electron cyclotron emission lines in the X-ray spectra of magnetized and transient X-ray pulsars during  
their low luminosity quiescent state (i.e  L$_x\leq$10$^{34}$ erg   s$^{-1}$). The energy line center  
is expected to peak at energies in the range $\sim$2--20 keV and it  should be superposed on the underlying 
soft thermal emission.  Emission-like features, similar to the ones predicted,  have been 
observed in the X-ray spectrum of a handful of transient X-ray pulsars in HMXB systems during low luminosity states (Nelson et al. 1995). 
We point out that the emission feature observed from IGR J18483$-$0311 could be similar to the one predicted by Nelson et al. (1995, 1993).
In our specific case, the lowest quiescent X-ray luminosity state achieved during the apastron passage 
as well as the sufficiently long {\itshape XMM-Newton} observation 
could have possibly favoured  the detection of a putative cyclotron emission line. This interpretation would imply a neutron star 
magnetic field value in the range (3--4)$\times$10$^{11}$ G, which is in very good agreement with
that independently inferred by us in the framework of accretion from  a
spherically symmetric stellar wind model for the source. It is also worth pointing out that the estimated 
magnetic field strength is of the order expected  for neutron stars known to display X-ray pulsations in high mass X-ray binaries, 
i.e. $\sim$(0.5--4)$\times$10$^{12}$ G. Consecutively, it is strong enough to affect the accretion flow within a few neutron star radii 
and cause the observed X-ray pulsations in IGR J18483$-$0311.

To date, IGR J18483$-$0311 is the only SFXT, 
among $\sim$ 10 firm members of this class, which exhibits  such intriguing  emission feature with important 
implications on theoretical models. However, 
this could be explained with the 
lack of sufficiently long X-ray 
observations of SFXTs at apastron passage (i.e. during their lowest quiescent X-ray states) with appropriate and sensitive enough 
X-ray facilities. 

Unfortunately,  we can go no further on these issues because the available exposure time and statistics of the data 
prevent us from a more detailed  investigation. Longer X-ray observations of  
IGR J18483$-$0311 using {\itshape XMM-Newton}, for example, are strongly  needed 
in order to achieve a  higher signal to noise. This would allow us  a much deeper investigation, 
in order to  support or reject our proposed interpretation in terms of electron cyclotron emission line.

\section*{Acknowledgments}
We thank an anonymous referee for useful comments which improved the quality of this paper.
The authors acknowledge the ASI financial support via grant ASI-INAF I/088/06/0/ and I/008/07/0/. 
VS is greatly indebted to Eleonora Torresi for valued suggestions, insightful comments and kind help with the XMM-Newton data analysis.

\end{document}